\documentclass[sigconf]{acmart}
\usepackage{multirow}
\usepackage{booktabs}
\usepackage[toc,page]{appendix}
\usepackage{amsmath}
\usepackage{pifont}
\usepackage{diagbox}
\usepackage{balance}
\AtBeginDocument{%
  }

\copyrightyear{2023}
\acmYear{2023}
\setcopyright{acmlicensed}\acmConference[KDD '23]{Proceedings of the 29th ACM SIGKDD Conference on Knowledge Discovery and Data Mining}{August 6--10, 2023}{Long Beach, CA, USA}
\acmBooktitle{Proceedings of the 29th ACM SIGKDD Conference on Knowledge Discovery and Data Mining (KDD '23), August 6--10, 2023, Long Beach, CA, USA}
\acmPrice{15.00}
\acmDOI{10.1145/3580305.3599918}
\acmISBN{979-8-4007-0103-0/23/08}

\begin{document}

\title{TransAct: Transformer-based Realtime User Action Model for Recommendation at Pinterest}



\author{Xue Xia}
\email{xxia@pinterest.com}
\affiliation{%
  \institution{Pinterest}
  \streetaddress{P.O. Box 1212}
  \city{San Francisco}
  \state{CA}
  \country{USA}
}

\author{Pong Eksombatchai}
\email{pong@pinterest.com}
\affiliation{%
  \institution{Pinterest}
  \streetaddress{P.O. Box 1212}
  \city{San Francisco}
  \state{CA}
  \country{USA}
}

\author{Nikil Pancha}
\email{npancha@pinterest.com}
\affiliation{%
  \institution{Pinterest}
  \streetaddress{P.O. Box 1212}
  \city{San Francisco}
  \state{CA}
  \country{USA}
}

\author{Dhruvil Deven Badani}
\email{dbadani@pinterest.com}
\affiliation{%
  \institution{Pinterest}
  \streetaddress{P.O. Box 1212}
  \city{San Francisco}
  \state{CA}
  \country{USA}
}

\author{Po-Wei Wang}
\email{poweiwang@pinterest.com}
\affiliation{%
  \institution{Pinterest}
  \streetaddress{P.O. Box 1212}
  \city{San Francisco}
  \state{CA}
  \country{USA}
}

\author{Neng Gu}
\email{ngu@pinterest.com}
\affiliation{%
  \institution{Pinterest}
  \streetaddress{P.O. Box 1212}
  \city{San Francisco}
  \state{CA}
  \country{USA}
}

\author{Saurabh Vishwas Joshi}
\email{sjoshi@pinterest.com}
\affiliation{%
  \institution{Pinterest}
  \streetaddress{P.O. Box 1212}
  \city{San Francisco}
  \state{CA}
  \country{USA}
}

\author{Nazanin Farahpour}
\email{nfarahpour@pinterest.com}
\affiliation{%
  \institution{Pinterest}
  \streetaddress{P.O. Box 1212}
  \city{San Francisco}
  \state{CA}
  \country{USA}
}

\author{Zhiyuan Zhang}
\email{zhiyuan@pinterest.com}
\affiliation{%
  \institution{Pinterest}
  \streetaddress{P.O. Box 1212}
  \city{San Francisco}
  \state{CA}
  \country{USA}
}

\author{Andrew Zhai}
\authornote{work done at Pinterest}
\email{andrew@aideate.ai}
\affiliation{%
  \institution{Pinterest}
  \streetaddress{P.O. Box 1212}
  \city{San Francisco}
  \state{CA}
  \country{USA}
}

\renewcommand{\shortauthors}{Xia et~al.}


\newcommand{\td}[1]{{\bf\color{red}[{\sc TODO:} #1]}}
\newcommand{\nikil}[1]{{\bf\color{red}[{\sc Nikil:} #1]}}
\newcommand{\xue}[1]{{\bf\color{red}[{\sc Xue:} #1]}}

\newcommand{\mA}{\boldsymbol{A}}
\newcommand{\mB}{\boldsymbol{B}}
\newcommand{\mC}{\boldsymbol{C}}
\newcommand{\mD}{\boldsymbol{D}}
\newcommand{\mE}{\boldsymbol{E}}
\newcommand{\mF}{\boldsymbol{F}}
\newcommand{\mH}{\boldsymbol{H}}
\newcommand{\mI}{\boldsymbol{I}}
\newcommand{\mO}{\boldsymbol{O}}
\newcommand{\mP}{\boldsymbol{P}}
\newcommand{\mR}{\boldsymbol{R}}
\newcommand{\mS}{\boldsymbol{S}}
\newcommand{\mU}{\boldsymbol{U}}
\newcommand{\mV}{\boldsymbol{V}}
\newcommand{\mW}{\boldsymbol{W}}
\newcommand{\mM}{\boldsymbol{M}}
\newcommand{\mX}{\boldsymbol{X}}
\newcommand{\mY}{\boldsymbol{Y}}

\newcommand{\mSigma}{\boldsymbol{\Sigma}}
\newcommand{\mLambda}{\boldsymbol{\Lambda}}
\newcommand{\mPsi}{\boldsymbol{\Psi}}
\newcommand{\mPhi}{\boldsymbol{\Phi}}
\newcommand{\matV}{\boldsymbol{\mathrm{V}}}
\newcommand{\matU}{\boldsymbol{\mathrm{U}}}
\newcommand{\matSigma}{\boldsymbol{\Sigma'}}

\newcommand{\va}{\boldsymbol{a}}
\newcommand{\vA}{\boldsymbol{A}}
\newcommand{\vb}{\boldsymbol{b}}
\newcommand{\vc}{\boldsymbol{c}}
\newcommand{\ve}{\boldsymbol{e}}
\newcommand{\vf}{\boldsymbol{f}}
\newcommand{\vg}{\boldsymbol{g}}
\newcommand{\vh}{\boldsymbol{h}}

\newcommand{\vS}{\boldsymbol{S}}
\newcommand{\vH}{\boldsymbol{H}}
\newcommand{\vl}{\boldsymbol{l}}
\newcommand{\vO}{\boldsymbol{O}}
\newcommand{\vo}{\boldsymbol{o}}
\newcommand{\vp}{\boldsymbol{p}}
\newcommand{\vq}{\boldsymbol{q}}
\newcommand{\vr}{\boldsymbol{r}}
\newcommand{\vu}{\boldsymbol{u}}
\newcommand{\vw}{\boldsymbol{w}}
\newcommand{\vx}{\boldsymbol{x}}
\newcommand{\vy}{\boldsymbol{y}}
\newcommand{\vz}{\boldsymbol{z}}
\newcommand{\vze}{\boldsymbol{0}}
\newcommand{\vone}{\boldsymbol{1}}

\newcommand{\vxi}{\boldsymbol{\xi}}
\newcommand{\vmu}{\boldsymbol{\mu}}
\newcommand{\vlambda}{\boldsymbol{\lambda}}
\newcommand{\vpsi}{\boldsymbol{\psi}}
\newcommand{\vphi}{\boldsymbol{\phi}}
\newcommand{\vsigma}{\boldsymbol{\sigma}}
\newcommand{\veps}{\boldsymbol{\varepsilon}}
\renewcommand{\epsilon}{\varepsilon}

\newcommand{\calA}{\mathcal{A}}
\newcommand{\calB}{\mathcal{B}}
\newcommand{\calD}{\mathcal{D}}
\newcommand{\calE}{\mathcal{E}}
\newcommand{\calL}{\mathcal{L}}
\newcommand{\calM}{\mathcal{M}}
\newcommand{\calN}{\mathcal{N}}
\newcommand{\calO}{\mathcal{O}}
\newcommand{\calP}{\mathcal{P}}
\newcommand{\calR}{\mathcal{R}}
\newcommand{\calT}{\mathcal{T}}
\newcommand{\calX}{\mathcal{X}}
\newcommand{\calY}{\mathcal{Y}}
\newcommand{\calZ}{\mathcal{Z}}





    

\newcommand{\tikzxmark}{%
\tikz[scale=0.23] {
    \draw[line width=0.7,line cap=round] (0,0) to [bend left=6] (1,1);
    \draw[line width=0.7,line cap=round] (0.2,0.95) to [bend right=3] (0.8,0.05);
}}
\newcommand{\tikzcmark}{%
\tikz[scale=0.23] {
    \draw[line width=0.7,line cap=round] (0.25,0) to [bend left=10] (1,1);
    \draw[line width=0.8,line cap=round] (0,0.35) to [bend right=1] (0.23,0);
}}

\begin{abstract}
Sequential models that encode user activity for next action prediction have become a popular design choice for building web-scale personalized recommendation systems. 
Traditional methods of sequential recommendation either utilize end-to-end learning on realtime user actions, or learn user representations separately in an offline batch-generated manner. 
This paper (1) presents Pinterest's ranking architecture for Homefeed, our personalized recommendation product and the largest engagement surface; 
(2) proposes TransAct, a sequential model that extracts users' short-term preferences from their realtime activities; 
(3) describes our hybrid approach to ranking, which combines end-to-end sequential modeling via TransAct with batch-generated user embeddings. 
The hybrid approach allows us to combine the advantages of responsiveness from learning directly on realtime user activity with the cost-effectiveness of batch user representations learned over a longer time period. 
We describe the results of ablation studies, the challenges we faced during productionization, and the outcome of an online A/B experiment, which validates the effectiveness of our hybrid ranking model. We further demonstrate the effectiveness of TransAct on other surfaces such as contextual recommendations and search.
Our model has been deployed to production in Homefeed, Related Pins, Notifications, and Search at Pinterest.



\end{abstract}



\maketitle

\section{Introduction}


The proliferation of online content in recent years has created an overwhelming amount of information for users to navigate. 
To address this issue, recommender systems are employed  in various industries to help users find relevant items from a vast selection, including products, images, videos, and music. 
By providing personalized recommendations, businesses and organizations can better serve their users and keep them engaged with the platform. 
Therefore, recommender systems are vital for businesses as they drive growth by boosting engagement, sales, and revenue.

\begin{figure}[!ht]
  \centering
  \includegraphics[width=0.85\linewidth]{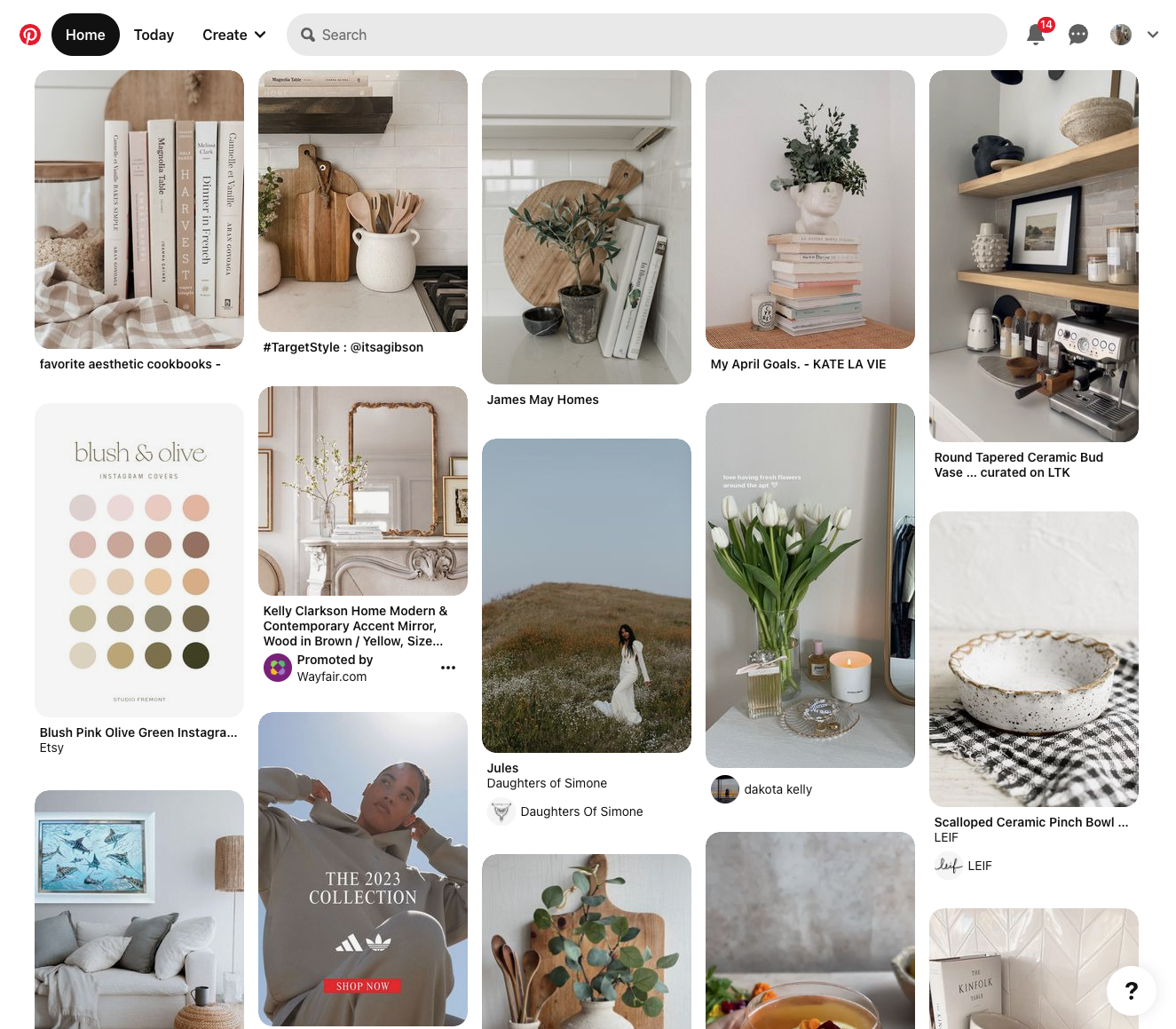}
  \caption{Pinterest Homefeed Page}
  \label{fig:hf}
  \Description{Pinterest Homefeed page}
\end{figure}

As one of the largest content sharing and social media platforms, Pinterest hosts billions of pins with rich contextual and visual information, and brings inspiration to over 400 million users. 
Upon visiting Pinterest, users are immediately presented with the Homefeed page as shown in Figure~\ref{fig:hf}, which serves as the primary source of inspiration and accounts for the majority of overall user engagement on the platform.
The Homefeed page is powered by a 3-stage recommender system that retrieves, ranks, and blends content based on user interests and activities. 
At the retrieval stage, we filter billions of pins created on Pinterest to thousands, based on a variety of factors such as user interests, followed boards, etc. Then we use a pointwise ranking model to rank candidate pins by predicting their personalized relevance to users. 
Finally, the ranked result is adjusted using a blending layer to meet business requirements.

Realtime recommendation is crucial because it provides a quick and up-to-date recommendation to users, improving their overall experience and satisfaction. 
The integration of realtime data, such as recent user actions, results in more accurate recommendations and increases the probability of users discovering relevant items~\cite{alibaba_seq_tfmr, pi2020search}.

Longer user action sequences result in improved user representation and hence better recommendation performance. 
However, using long sequences in ranking poses challenges to infrastructure, as they require significant computational resources and can result in increased latency.
To address this challenge, some approaches have utilized hashing and nearest neighbor search in long user sequences~\cite{pi2020search}.
Other work encodes users' past actions over an extended time frame to a user embedding~\cite{pinnerformer} to represent long-term user interests. User embedding features are often generated as \textit{batch} features (e.g. generated daily), which are cost-effective to serve across multiple applications with low latency. 
The limitation of existing sequential recommendation is that they either only use realtime user actions, or only use a batch user representation learned from long-term user action history.

We introduce a novel realtime-batch hybrid ranking approach that combines both \textit{realtime} user action signals and \textit{batch} user representations. 
To capture the realtime actions of users, we present TransAct - a new transformer-based module designed to encode recent user action sequences and comprehend users' immediate preferences.
For user actions that occur over an extended period of time, we transform them into a batch user representation~\cite{pinnerformer}.

By combining the expressive power of TransAct with batch user embeddings, the hybrid ranking model offers users realtime feedback on their recent actions, while also accounting for their long-term interests. The realtime component and batch component complement each other for recommendation accuracy. This leads to an overall improvement in the user experience on the Homefeed page.

The major contributions of this paper are summarized as follows:
\begin{itemize}
 \item We describe Pinnability, the architecture of Pinterest's Homefeed production ranking system. The Homefeed personalized recommendation product accounts for the majority of the overall user engagement on Pinterest. 

 \item  We propose TransAct, a transformer-based realtime user action sequential model that effectively captures users' short-term interests from their recent actions. We demonstrate that combining TransAct with daily-generated user representations~\cite{pinnerformer} to a hybrid model leads to the best performance in Pinnability. This design choice is justified through a comprehensive ablation study. Our code implementation is publicly available\footnote{Our code is available on Github: \url{https://github.com/pinterest/transformer_user_action}}.

\item We describe the serving optimization implemented in Pinnability to make feasible the computational complexity increase of 65 times when introducing TransAct to the Pinnability model. Specifically, optimizations are done to enable GPU serving of our prior CPU-based model.
\item We describe online A/B experiments on a real-world recommendation system using TransAct. We demonstrate some practical issues in the online environment, such as recommendation diversity drop and engagement decay, and propose solutions to address these issues.  

\end{itemize}

The remainder of this paper is organized as follows: Related
work is reviewed in Section~\ref{sec:related_work}. Section~\ref{sec:method} describes the design of TransAct and the details of bringing it to production. Experiment results are reported in Section~\ref{sec:exp}. We discuss some findings beyond experiments in Section~\ref{sec:discussion}. Finally, we conclude our work in Section~\ref{sec:conclusion}.



\section{Related Work}
\label{sec:related_work}


\subsection{Recommender System}
Collaborative filtering (CF)~\cite{sarwar2001item, CF, CF2} makes recommendations based on the assumption that a user will prefer an item that other similar users prefer. 
It uses the user behavior history to compute the similarity between users and items and recommend items based on similarity. 
This approach suffers from the sparsity of the user-item matrix and cannot handle users who have never interacted with any items. 
Factorization machines~\cite{FM, rendle2010factorization}, on the other hand, are able to handle sparse matrices.

More recently, deep learning (DL) has been used in click-through rate (CTR) prediction tasks. 
For example, Google uses Wide \& Deep~\cite{cheng2016wide} models for application recommendation. 
The wide component achieves memorization by capturing the interaction between features, while the deep component helps with generalization by learning the embedding of categorical features using a feed forward network.
DeepFM~\cite{guo2017deepfm} makes improvements by learning both low-order and high-order feature interactions automatically. 
DCN~\cite{dcn} and its upgraded version DCN v2~\cite{DCNv2} both aim to automatically model the explicit feature crosses.
The aforementioned recommender systems do not work well in capturing the short-term interests of users since only the static features of users are utilized. 
These methods also tend to ignore the sequential relationship within the action history of a user, resulting in an inadequate representation of user preferences. 


\subsection{Sequential Recommendation}
To address this problem, sequential recommendation has been widely studied in both academia and the industry. 
A sequential recommendation system uses a behavior history of users as input and applies recommendation algorithms to suggest appropriate items to users.
Sequential recommendation models are able to capture users' long-term preferences over an extended period of time, similar to traditional recommendation methods. Additionally, they also have the added benefit of being able to account for users' evolving interests, which enables higher quality recommendations.

Sequential recommendation is often viewed as a next item prediction task, where the goal is to predict a user's next action based on their past action sequence. 
We are inspired by the previous sequential recommendation method \cite{alibaba_seq_tfmr} in terms of encoding users' past action into a dense representation. 
Some early sequential recommendation systems use machine learning techniques, such as Markov Chain~\cite{he2016fusingMC} and session-based K nearest neighbors (KNN)~\cite{hu2020modelingknn} to model the temporal dependencies among interactions in users' action history. 
These models are criticized for not being able to fully capture the long-term patterns of users by simply combining information from different sessions. 
Recently, deep learning techniques such as recurrent neural networks (RNN)~\cite{rnn} have shown great success in natural language processing and have become increasingly popular in sequential recommendation. 
As a result, many DL-based sequential models~\cite{donkers2017sequential, hidasi2015session, tan2016improved, zhou2019deep} have achieved outstanding performance using RNNs. 
Convolutional neural networks (CNNs)~\cite{cnn} are widely used for processing time-series data and image data. 
In the context of sequential recommendation, CNN-based models can effectively learn dependency within a set of items users recently interacted with, and make recommendations accordingly~\cite{tang2018personalized, tuan20173d}.
Attention mechanism is originated from the neural machine translation task, which models the importance of different parts of the input sentences on the output words~\cite{bahdanau2014neural}.
Self-attention is a mechanism known to weigh the importance of different parts of an input sequence~\cite{tfmr}. 
There have been more recommender systems that use attention~\cite{DIN} and self-attention~\cite{zhang2019next, alibaba_seq_tfmr, li2020time, SASRec, sun2019bert4rec}. 

Many previous works~\cite{SASRec,li2020time,sun2019bert4rec}
only perform offline evaluations using public datasets. However, the online environment is more challenging and unpredictable. Our method is not directly comparable to these works due to differences in the problem formulation. Our approach resembles a Click-through Rate (CTR) prediction task.
Deep Interest Network (DIN) uses an attention mechanism to model the dependency within users' past actions in CTR prediction tasks. 
Alibaba's Behavior Sequence Transformer (BST)~\cite{alibaba_seq_tfmr} is the improved version of DIN and is closely related to our work.
They propose to use Transformer to capture the user interest from user actions, emphasizing the importance of the action order.
However, we found that positional information does not add much value. We find other designs like better early fusion and action type embedding are effective when dealing with sequence features.



\section{Methodology}
\label{sec:method}

In this section, we introduce TransAct, our realtime-batch hybrid ranking model. 
We will start with an overview of the Pinterest Homefeed ranking model, Pinnability. We then describe how to use TrancAct to encode the realtime user action sequence features in Pinnability for the ranking task.

\subsection{Preliminary: Homefeed Ranking Model}
\label{sec:pinnability}
In Homefeed ranking, we model the recommendation task as a pointwise multi-task prediction problem, which can be defined as follows: given a user $u$ and a pin $p$, we build a function to predict the probabilities of user $u$ performing different actions 
on the candidate pin $p$. The set of different actions contains both positive and negative actions, e.g. click, repin\footnote{A "repin" on Pinterest refers to the action of saving an existing pin to another board by a user.} and hide.

We build \textit{Pinnability}, Pinterest's Homefeed ranking model, to approach the above problem.
The high-level architecture is a Wide and Deep learning (WDL) model~\cite{cheng2016wide}.  
The Pinnability model utilizes various types of input signals, such as user signals, pin signals, and context signals.
These inputs can come in different formats, including categorical, numerical, and embedding features.

\begin{figure}[h]
  \centering
  \includegraphics[width=0.9\linewidth]{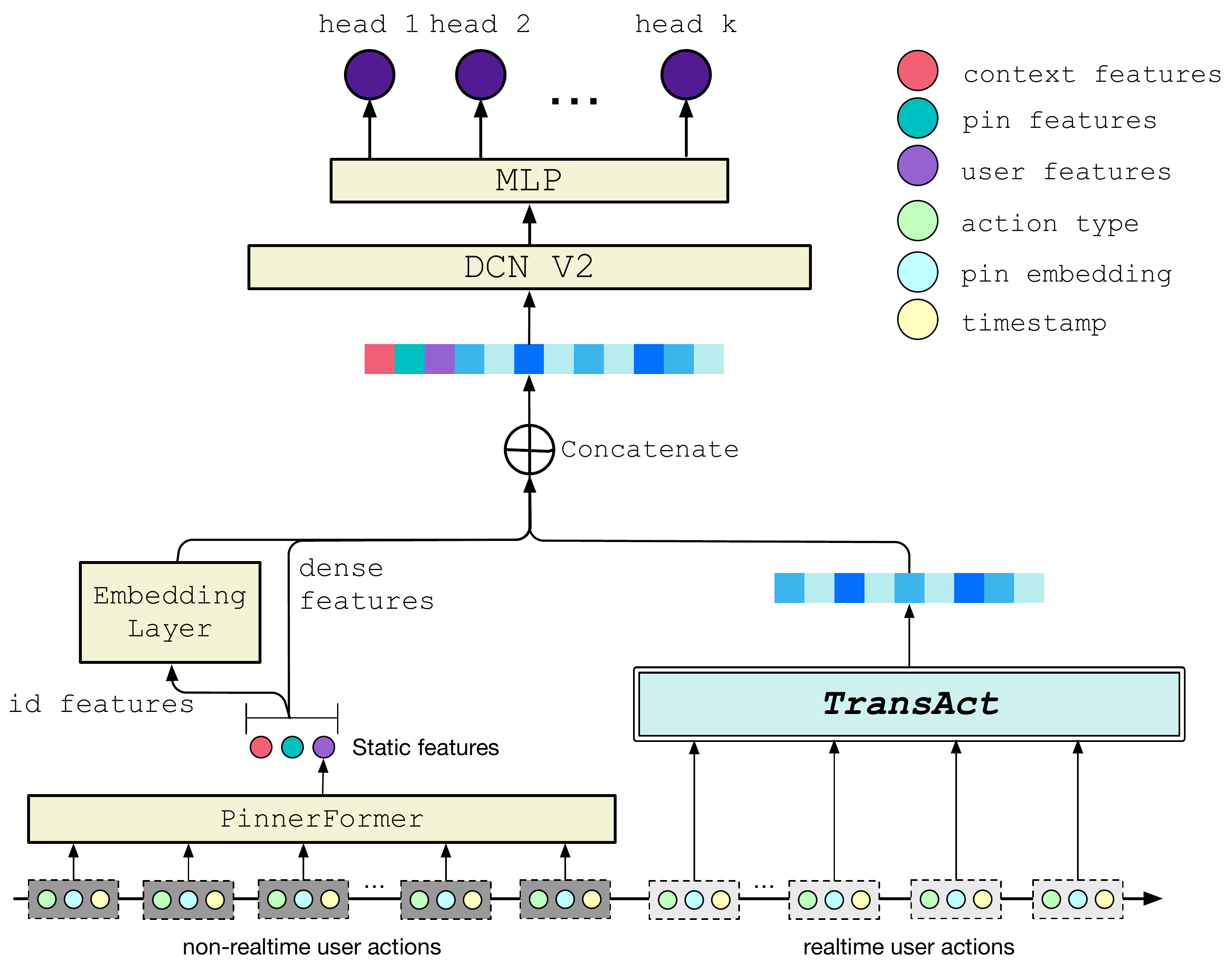}
  \caption{Pinterest Homefeed ranking model (Pinnability)}
  \label{fig:pinnability}
  \Description{Pinterest Homefeed ranking model (Pinnability)}
\end{figure}

We use embedding layers to project categorical features to dense features, and perform batch normalization on numerical features. We then apply a feature cross using a full-rank DCN V2~\cite{DCNv2} to explicitly model feature interactions.
At last, we use fully connected layers with a set of output action heads $\vH = \{h_1, h_2, \dots , h_k\}$ to predict the user actions on the candidate pin $p$. Each head maps to one action.
As shown in Figure~\ref{fig:pinnability}, our model is a realtime-batch hybrid model that encodes the user action history features by both realtime (TransAct) and batch (PinnerFormer) approaches and optimizes for the ranking task~\cite{10.1145/3523227.3547394}.


Each training sample is $(\vx, \vy)$, where $\vx$ represents a set of features, and $\vy \in \{0,1\} ^{|\vH|}$. Each entry in $\vy$ corresponds to the label of an action head in $\vH$.
The loss function of Pinnability is a weighted cross-entropy loss, designed to optimize for multi-label classification tasks.
We formulate the loss function as:

\begin{equation}
\label{eq:loss}
\mathcal{L} = w_u \sum_{h \in H } \left\{ - w_h\left[y_h \log f(\vx)_h + (1-y_h)(1 - \log f(\vx)_h) \right] \right\}
\end{equation}
where $f(\vx)\in (0, 1)^H$, and $f(\vx)_h$ is the output probability of head $h$. $y_h \in \{0,1\}$ is the ground truth on head $h$.

A weight $w_h$ is applied on the cross entropy of each head's output $f(\vx)_h$.
$w_h$ is calculated using the ground truth $\vy$ and a label weight matrix  $\mM \in \mathbb{R} ^{|H| * |H|}$ as follows:

\begin{equation}
w_h = \sum_{a \in H }\mM_{h,a} \times y_a
\end{equation}

The label weight matrix $\mM$ acts as a controlling factor for the contribution of each action to the loss term of each head\footnote{For more details, see Appendix ~\ref{appendix:head_weights}}.
Note that if $\mM$ is a diagonal matrix, Eq~\eqref{eq:loss} reduces to a standard multi-head binary cross entropy loss.
But selecting empirically determined label weights $\mM$ improves performance considerably.

In addition, each training example is weighted by a user-dependent weight $w_u$, which is determined by user attributes, such as the user state\footnote{User states are used to group users of different behavior patterns, for example, users who engage daily are in one group, while those who engage once a month have a different user state}, gender and location.
We compute $w_u$ by multiplying user state weight, user gender weight, and user location weight: $w_u = w_{\text{state}} \times w_{\text{location}} \times w_{\text{gender}}$. These weights are adjusted based on specific business needs.


\subsection{Realtime User Action Sequence Features}

User's past action history is naturally a variable length feature -- different users have different amounts of past actions on the platform. 
  
Although a longer user action sequence usually means more accurate user interest representation, in practice, it is infeasible to include all user actions. Because the time needed to fetch user action features and perform ranking model inference can also grow substantially, which in turn hurts user experience and system efficiency. Considering infrastructure cost and latency requirements, we choose to include each user's most recent 100 actions in the sequence. For users with less than 100 actions, we pad the feature to the length of 100 with 0s. The user action sequence features are sorted by timestamp in descending order, i.e. the first entry being the most recent action.

All actions in the user action sequence are pin-level actions. For each action, we use three primary features: the timestamp of the action, action type, and the 32-dimensional PinSage embedding \cite{PinSage} of the pin. 
PinSage is a compact embedding that encodes a pin's content information. 

\subsection{Our Approach: TransAct}

Unlike static features, the realtime user action sequence feature $\vS(u) = [a_1,a_2, ...,a_n]$ is handled using a specialized sub-module called TransAct. TransAct extracts sequential patterns from the user's historical behavior and predicts $(u, p)$ relevance scores.

\subsubsection{Feature encoding}
The relevance of pins that a user has engaged with can be determined by the types of actions taken on them in the user's action history. For example, a pin repinned to a user's board is typically considered more relevant than one that the user only viewed. If a pin is hidden by the user, the relevance should be very low. To incorporate this important information, we use trainable embedding tables to project action types to low-dimensional vectors. The user action type sequence is then projected to a user action embedding matrix $\mW_{actions} \in \mathbb{R} ^{|S| \times d_{action}}$, where $d_{action}$ is the dimension of action type embedding.

As mentioned earlier, the content of pins in the user action sequence is represented by PinSage embeddings \cite{PinSage}. Therefore, the content of all pins in the user action sequence is a matrix $\mW_{pins} \in \mathbb{R} ^{|S| \times d_{PinSage}}$. 
The final encoded user action sequence feature is \texttt{CONCAT}$(\mW_{actions},\mW_{pins}) \in \mathbb{R} ^{|S| \times (d_{PinSage}+d_{action})}$.

\subsubsection{Early fusion}
\label{sec:earlyfusion}
One of the unique advantages of using  user action sequence features directly in the ranking model is that we can explicitly model the interactions between the candidate pin and the user's engaged pins. 
Early fusion in recommendation tasks refers to merging user and item features at an early stage of the recommendation model. 
Through experiments, we find that early fusion is an important factor to improve ranking performance. Two early fusion methods are evaluated:
\begin{itemize}
\item \texttt{append}: Append candidate pin's PinSage embedding to user action sequence as the last entry of the sequence, similar to BST~\cite{alibaba_seq_tfmr}. Use a zero vector to serve as a dummy action type for candidate pin.
\item \texttt{concat}: For each action in the user action sequence, concatenate the candidate pin's PinSage embedding with user action features.
\end{itemize}
We choose  \texttt{concat}  as our early fusion method based on the offline experiment results. The resulting sequence feature with early fusion is a 2-d matrix $\mU \in \mathbb{R} ^{|S| \times d}$, where $d = (d_{action} + 2d_{PinSage})$

\subsubsection{Sequence Aggregation Model}
\label{sec: seq_model}
With the user action sequence feature $\mU$ prepared, the next challenge is to efficiently aggregate all the information in the user action sequence to represent the user's short-term preference. 
Some popular model architectures for sequential modeling in the industry include CNN\cite{cnn}, RNN~\cite{rnn} and recently transformer~\cite{tfmr}, etc. 
We experimented with different sequence aggregation architectures and choose transformer-based architectures. We employed the standard transformer encoder with 2 encoder layers and one head. The hidden dimension of feed forward network is denoted as $d_{hidden}$. Positional encoding is not used here because our offline experiment showed that position information is ineffective\footnote{For more details about positional encoding, see Appendix ~\ref{appendix:pe}.}.

\subsubsection{Random Time Window Mask}
Training on all recent actions of a user can lead to a rabbit hole effect, where the model recommends content similar to the user's recent engagements.
This hurts the diversity of users' Homefeeds, which is harmful to long-term user retention.
To address this issue, we use the timestamps of the user action sequence to build a time window mask for the transformer encoder. 
This mask filters out certain positions in the input sequence before the self-attention mechanism is applied. 
In each forward pass, a random time window $T$ is sampled uniformly from 0 to 24 hours. All actions taken within $(t_{request} - T, t_{request})$ are masked, where $t_{request}$ stands for the timestamp of receiving the ranking request. It is important to note that the random time window mask is only applied during training, while at inference time, the mask is not used. 

\subsubsection{Transformer Output Compression}
The output of the transformer encoder is a matrix $\mO = (\vo_0:\vo_{|S|-1})\in \mathbb{R} ^ {|S| \times d}$. We only take the first $K$ columns $(\vo_0:\vo_{K-1})$, concatenated them with the max pooling vector \texttt{MAXPOOL}$(\mO) \in \mathbb{R}^{d}$, and flattened it to a vector $\vz \in \mathbb{R} ^{(K + 1) * d}$. The first $K$ output columns capture  users’ most recent interests and \texttt{MAXPOOL}$(\mO)$ represents users' longer-term preference over $S(u)$. Since the output is compact enough, it can be easily integrated into the Pinnability framework using the DCN v2~\cite{DCNv2}  feature crossing layer.

\begin{figure}[h]
  \centering
  \includegraphics[width=0.9\linewidth]{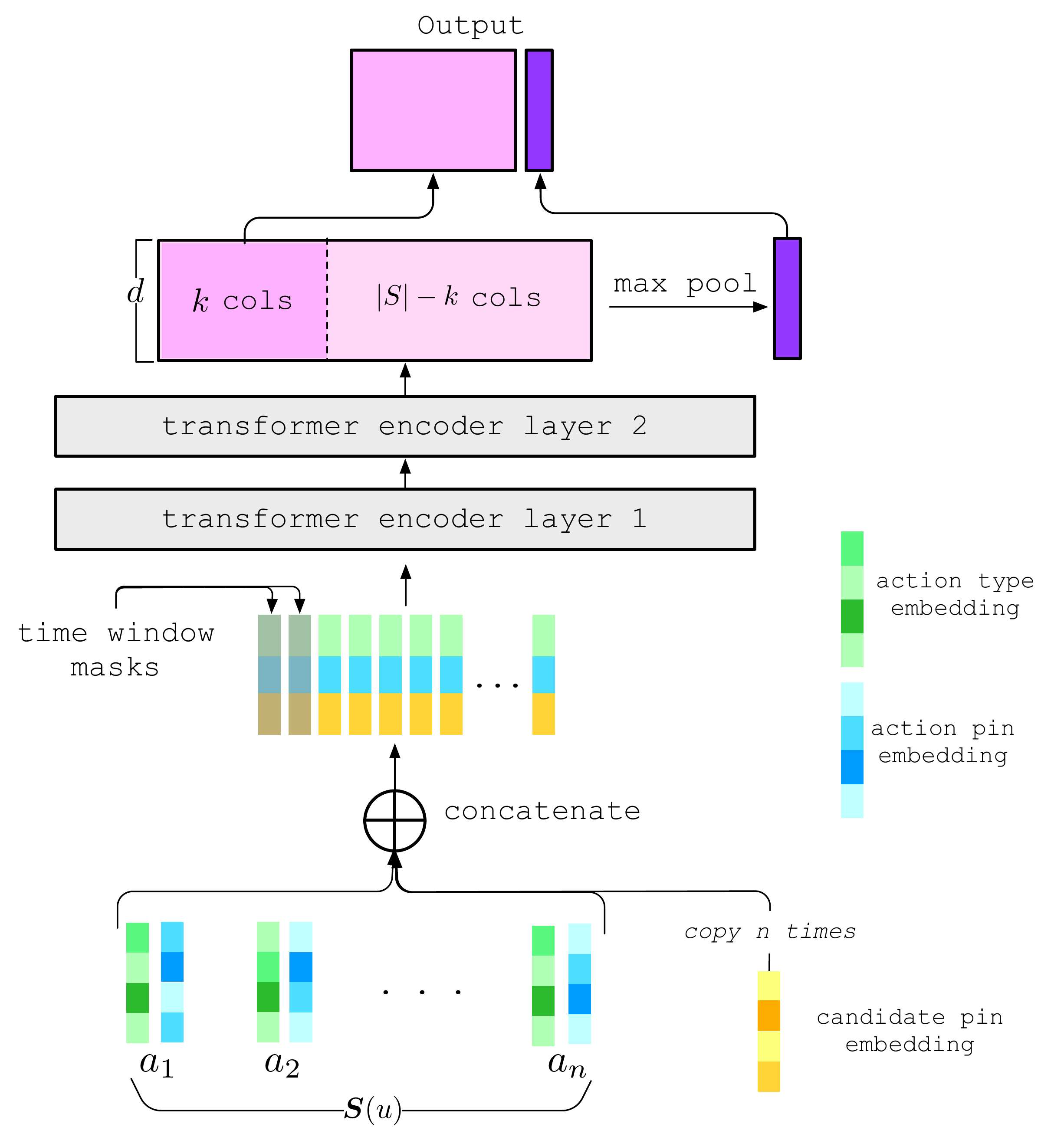}
  \caption{TransAct architecture. Note that this is a submodule that can be plugged into any similar architecture like Pinnability}
  \label{fig:seq_encoder}
  \Description{TransAct model architecture}
\end{figure}

\subsection{Model Productionization}

\subsubsection{Model Retraining}
Retraining is important for recommender systems because it allows the system to continuously adapt to changing user behavior and preferences over time. Without retraining, a recommender system's performance can degrade as the user's behavior and preferences change, leading to less accurate recommendations~\cite{tech_debt}. 
This holds especially true when we use realtime features in ranking. The model is more time sensitive and requires frequent retraining. 
Otherwise, the model can become stale in a matter of days, leading to less accurate predictions. 
We retrain Pinnability from scratch twice per week. We find that this retraining frequency is essential to ensure a consistent engagement rate and still maintain a manageable training cost.
We will dive into the importance of retraining in Section~\ref{sec:retrain_exp}.

\subsubsection{GPU serving} Pinnability with TransAct is 65 times more computationally complex compared to its predecessors in terms of floating point operations. Without any breakthroughs in model inference, our model serving cost and latency would increase by the same scale. GPU model inference allows us to serve Pinnability with TransAct at neutral latency and cost\footnote{For more details about model effiency, see Appendix ~\ref{appendix:efficiency}.}. 

The main challenge to serve Pinnability on GPUs is the CUDA kernel launch overhead. The CPU cost of launching operations on the GPU is very high, but it is often overshadowed by the prolonged GPU computation time. However, this is problematic for Pinnability GPU model serving in two ways. First, Pinnability and recommender models in general process hundreds of features, which means that there is a large number of CUDA kernels. Second, the batch size during online serving is small and hence each CUDA kernel requires little computation. With a large number of small CUDA kernels, the launching overhead is much more expensive than the actual computation. We solved the technical challenge through the following optimizations:

\textbf{Fuse CUDA kernels.} An effective approach is to fuse operations as much as possible. We leverage standard deep learning compilers such as nvFuser\footnote{\url{https://pytorch.org/blog/introducing-nvfuser-a-deep-learning-compiler-for-pytorch/}} but often found human intervention is needed for many of the remaining operations. One example is our embedding table lookup module, which consists of two computation steps: raw id to table index lookup and table index to embedding lookup. This is repeated hundreds of times due to the large number of features. We significantly reduce the number of operations by leveraging cuCollections\footnote{\url{https://github.com/NVIDIA/cuCollections}} to support hash tables for the raw ids on GPUs and implementing a custom consolidated embedding lookup module to merge the lookup for multiple features into one lookup. As a result, we reduced hundreds of operations related to sparse features into one.

\textbf{Combine memory copies.} For every inference, hundreds of features are copied from the CPU to the GPU memory as individual tensors. The overhead of scheduling hundreds of tensor copies becomes the bottleneck. To decrease the number of tensor copy operations, we combine multiple tensors into one continuous buffer before transferring them from CPU to GPU. This approach reduces the scheduling overhead of transferring hundreds of tensors individually to transferring one tensor. 

\textbf{Form larger batches.} For CPU-based inference, smaller batches are preferred to increase parallelism and reduce latency. However, for GPU-based inference, larger batches are more efficient~\cite{sze2017efficient}. This led us to re-evaluate our distributed system setup. Initially, we used a scatter-gather architecture to split requests into small batches and run them in parallel on multiple leaf nodes for better latency. However, this setup did not work well with GPU-based inference. Instead, we use the larger batches in the original requests directly. To compensate for the loss of cache capacity, we implemented a hybrid cache that uses both DRAM and SSD.

\textbf{Utilize CUDA graphs.} We relied on CUDA Graphs\footnote{\url{https://developer.nvidia.com/blog/cuda-graphs/}} to completely eliminate the remaining small operations overhead. CUDA Graphs capture the model inference process as a static graph of operations instead of individually scheduled ones, allowing the computation to be executed as a single unit without any kernel launching overheads. 

\subsubsection{Realtime Feature Processing}

When a user takes an action, a realtime feature processing application based on Flink\footnote{https://flink.apache.org/}
 consumes user action Kafka\footnote{https://kafka.apache.org/}
 streams generated from front-end events. It validates each action record, detects and combines duplicates, and manages any time discrepancies from multiple data sources. The application then materializes the features and stores them in Rockstore~\cite{rockstore_blog}. At serving time, each Homefeed logging/serving request triggers the processor to convert sequence features into a format that can be utilized by the model.
 
\section{Experiment}
\label{sec:exp}
In this section, we will present extensive offline and online A/B experiment results of TransAct. We compare TransAct with baseline models using Pinterest's internal training data. 
\subsection{Experiment Setup}
\subsubsection{Dataset}
We construct the offline training dataset from three weeks of Pinterest Homefeed view log (FVL). The model is trained on the first two weeks of FVL and evaluated on the third week. The training data is sampled based on user state and labels. 
For example, we design the sampling ratio for different label actions based on their statistical distribution and importance. 
In addition, since users only engage with a small portion of pins shown on their Homefeed page, most of the training samples are negative samples. To balance the highly skewed dataset and improve model accuracy,
we employ downsampling on the negative samples and set a fixed ratio between the positive and negative samples.
Our training dataset contains 3 billion training instances of 177 million users and 720 million pins.

In this paper, we conduct all experiments with the Pinterest dataset. We do not use public datasets as they lack the necessary realtime user action sequence metadata features, such as item embeddings and action types, required by TransAct.  
Furthermore, they are incompatible with our proposal of realtime-batch hybrid model, which requires both realtime and batch user features. 
And they cannot be tested in online A/B experiments. 

\subsubsection{Hyperparameters}
Realtime user sequence length is $|S| = 100$ and the dimension of action embedding $d_{action} = 32$. 
The encoded sequence feature is passed through a transformer encoder composed of 2 transformer blocks, with a default dropout rate of 0.1. 
The feed forward network in the transformer encoder layer has a dimension of $d_{hidden}=32$, and positional encoding is not used. 
The implementation is done using PyTorch. We use an Adam~\cite{adam} optimizer with a learning rate scheduler. The learning rate begins with a warm-up phase of 5000 steps, gradually increasing to 0.0048, and finally reduced through cosine annealing. The batch size is 12000.

\subsection{Offline Experiment}

\subsubsection{Metrics}

The offline evaluation data, unlike training data, is randomly sampled from FVL to represent the true distribution of the real-world traffic. With this sampling strategy, the offline evaluation data is representative of the entire population, reducing the variance of evaluation results.

In addition to sampling bias, we also eliminate position bias in offline evaluation data.
Position bias refers to the tendency for items at the top of a recommendation to receive more attention and engagement than the items lower down the list. This can be a problem when evaluating a ranking model, as it can distort the evaluation results and make it difficult to accurately assess the model's performance. To avoid position bias, we randomize the order of pins in a very small portion of Homefeed recommendation sessions. This is done by shuffling the recommendations before presenting them to users. We gather the FVL for those randomized sessions and only use randomized data to perform the offline evaluation.

Our model is evaluated on HIT@3.
A chunk $\vc = [p_1, p_2, \dots, p_n]$ refers to a group of pins that are recommended to a user at the same time. 
Each input instance to the ranking model is associated with a user id $u\_{id}$, a pin id $p\_{id}$, and a chunk id $c\_{id}$. 
The evaluation output is grouped by $(u\_{id}, c\_{id})$ so that it contains the model output from the same ranking request. We sort the pins from the same ranking request by a final ranking score $\mathcal{S}$, which is a linear combination of Pinnability output heads $f(\vx)$. 
\begin{equation}
\mathcal{S} = \sum_{h \in H } u_h f(\vx)_h
\end{equation}
Then we take the top $K$ ranked pins in each chunk and calculate the hit@K for all heads, denoted by $\beta_{c,h}$, which is defined as the number of topK-ranked pins whose labels of $h$ are 1. For example, if a chunk $\vc = [p_1, p_2, p_3,\dots , p_n]$ is sorted by $\mathcal{S}$, and the user repins $p_1$ and $p_4$, then hit@K of repin $\beta_{c,repin}=1$ when $K=3$. 

We calculate the aggregated HIT@3  for each head $h$ as follows:

\begin{equation}
HIT@3/h = \frac{\sum_{u \in U }\sum_{c \in C_u } \beta_{c,h}}{\left| U \right|}
\end{equation}

It is important to note that for actions indicating positive engagement, such as repin or click, a higher HIT@K score means better model performance. Conversely, for actions indicating negative engagement, such as hide, a lower HIT@K/hide score is desirable.

At Pinterest, a non-core user is defined as a user who has not actively saved pins to boards within the past 28 days.
Non-core users tend to be less active and therefore pose a challenge in terms of improving their recommendation relevance due to their limited historical engagement. This is also referred to as the cold-start user problem in recommendation~\cite{natarajan2020resolving}. Despite the challenges, it is important to retain non-core users as they play a crucial role in maintaining a diverse and thriving community, contributing to long-term platform growth.

All reported results are statistically significant (p-value $ < 0.05$) unless stated otherwise.
\subsubsection{Results}

\begin{table}
\caption{Offline evaluation of comparing existing methods with TransAct. ($^*$ statistically insignificant)} \label{tab:offline_res}
  \resizebox{.45\textwidth}{!}{%
  \begin{tabular}{ccccc}
    \toprule
    \multirow{2}{*}{Methods} & \multicolumn{2}{c}{HIT@3/repin}& \multicolumn{2}{c}{HIT@3/hide}\\ \cmidrule{2-5}%
       & all  & non-core  & all & non-core  \\
    \midrule
    WDL + seq & +0.21\% & +0.35\%& -1.61\% &  -1.55\% \\
    BST (all actions) & +4.41\% &+5.09\% &  +2.33\% &+3.59\% \\
    BST (positive actions) & +7.34\% &+8.16\% &  -1.12\%$^*$ & -3.14\%$^*$ \\
    TransAct & \textbf{+9.40\%} & \textbf{+10.42\%} & \textbf{-14.86}\% & \textbf{-13.54\%}\\
  \bottomrule
\end{tabular}%
}
\end{table}

We compare TransAct with existing methods of sequential recommendation. 
The first baseline is the WDL model~\cite{cheng2016wide} that incorporates sequence features as part of its wide features. Due to the large size of the sequence features, the number of parameters in the feature cross layer would grow quadratically, making it unfeasible for both training and online serving. 
Therefore, we used an averaging pooling for PinSage embeddings of user actions to encode the sequence.
The second baseline is Alibaba's behavior sequence transformer (BST) model~\cite{alibaba_seq_tfmr}. 
We trained 2 BST model variants here: one with only positive actions in user sequence, the other with all actions. 
We opted not to compare our results with DIN~\cite{DIN} as BST has already demonstrated its superiority over DIN. Additionally, we did not compare with variants like BERT4Rec~\cite{BERT4Rec} as the problem formulations are different and a direct comparison is not feasible.

The results of the model comparison are presented in Table~\ref{tab:offline_res}. It is evident that BST and TransAct outperform the WDL model, demonstrating the necessity of using a specialized sequential model to effectively capture short-term user preferences through real-time user action sequence features. 
BST performs well when only positive actions are encoded, however, it struggles to distinguish negative actions. 
In contrast, TransAct outperforms BST, particularly in terms of hide prediction, due to its ability to distinguish between different actions by encoding action types. 
Furthermore, TransAct also exhibits improved performance in HIT@3/repin compared to BST, which can be attributed to its effective early fusion and output compression design. 
A common trend across all groups is that the performance for non-core users is better than for all users, this is due to realtime user action features being crucial for users with limited engagement history on the platform, as they provide the only source of information for the model to learn their preferences.
\subsection{Ablation Study}

\subsubsection{Hybrid ranking model}
First, we investigate the effect of the realtime-batch hybrid design by examining the individual impact of TransAct(realtime component) and Pinnerformer(batch component). 
Table~\ref{tab:hybrid} shows the relative decrease in offline performance from the model containing all user features as we remove each component.
TransAct captures users' immediate interests, which contribute the most to the user's overall engagement, while PinnerFormer (PF)~\cite{pinnerformer} extracts users' long-term preferences from their historical behavior.
We observe that TransAct is the most important user understanding feature in the model, but we still see value from the large-scale training and longer-term interests captured by PinnerFormer, showing that longer-term batch user understanding can complement a realtime engagement sequence for recommendations.
In the last row of Table~\ref{tab:hybrid}, we show that removing all user features other than TransAct and PinnerFormer only leads to a relatively small drop in performance, demonstrating the effectiveness of our combination of a realtime sequence model with a pre-trained batch model.
\
\begin{table}
\caption{Ablation study of realtime-batch hybrid model}
  \label{tab:hybrid}
\resizebox{.4\textwidth}{!}{%
  \begin{tabular}{ccccc}
    \toprule
    \begin{tabular}{@{}c@{}}TransAct\end{tabular} & PF & \begin{tabular}{@{}c@{}}Other User\\Features\end{tabular} & HIT@3/repin & HIT@3/hide \\
    \midrule
    \checkmark & \checkmark & \checkmark  & --- & --- \\
    \checkmark & \ding{53} & \checkmark  & -2.46\% & +3.61\% \\
    \ding{53} & \checkmark & \checkmark  & -8.59\% & +17.45\% \\
    \checkmark & \checkmark & \ding{53}  & -0.67\% & +1.40\% \\
  \bottomrule
\end{tabular}%
}
\end{table}

\subsubsection{Base sequence encoder architecture}

We perform an offline evaluation on different sequential models that process realtime user sequence features. We use different architectures to encode the PinSage embedding sequence from users' realtime actions.

\textbf{Average Pooling}: use the average of PinSage embeddings in user sequence to present the user’s short-term interest

\textbf{CNN}: use a 1-d CNN with 256 output channels to encode the sequence. Kernel size is 4 and stride is 1. 

\textbf{RNN}: use 2 RNN layers with a hidden dimension of 256, to encode a sequence of PinSage embeddings. 

\textbf{LSTM}: use Long Short-Term Memory (LSTM)~\cite{lstm}, a more sophisticated version of RNN that better captures longer-term dependencies by using memory cells and gating. We use 2 LSTM layers with the hidden size of 256.

\textbf{Vanilla Transformer}: encodes only PinSage embeddings sequence directly using the Transformer encoder module. We use 2 transformer encoder layers with a hidden dimension of 32.
 

The baseline group is the Pinnability model without realtime user sequence feature.
From Table~\ref{tab:seq_encoder}, we learned that using realtime user sequence features, even with a simple average pooling method, improves engagement. 
Surprisingly, more complex architectures like RNN, CNN, and LSTM do not always perform better than average pooling. However, the best performance is achieved with the use of a vanilla transformer, as it significantly reduces HIT@3/hide and improves HIT@3/repin.

\begin{table}
  \caption{Offline evaluation of sequence encoder architecture}
  \label{tab:seq_encoder}
  \begin{tabular}{crr}
    \toprule
    Sequence Encoder & HIT@3/repin & HIT@3/hide\\
    \midrule
    Average Pooling & +0.21\%& -1.61\% \\
    CNN & +0.08\% & -1.29\% \\
    RNN & -1.05\% & -2.46\% \\
    LSTM & -0.75\% & -2.98\% \\
    Vanilla Transformer & \textbf{+1.56\%} & \textbf{-8.45\%} \\
  \bottomrule
\end{tabular}
\end{table}


\subsubsection{Early fusion and sequence length selection}

As discussed in Section~\ref{sec:earlyfusion}, early fusion plays a crucial role in the ranking model.  
By incorporating early fusion, the model can not only take into account the dependency between different items in the user's action history but also explicitly learn the relationship between the ranking candidate pin and each pin that the user has engaged with in the past.


Longer user action sequences naturally are more expressive than short sequences. 
To learn the effect of input sequence length on the model performance, we evaluate the model on different lengths of user sequence input.

\begin{figure}[h]
  \centering
  \includegraphics[width=\linewidth]{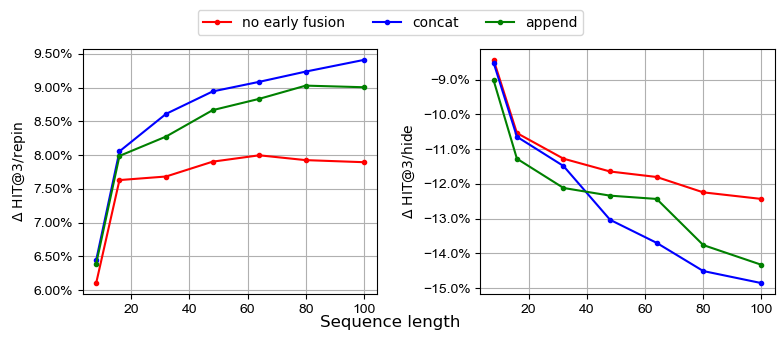}
  \caption{Effect of early fusion and sequence length on ranking model performance (HIT@3/repin, HIT@3/hide) }
  \label{fig:seq_len}
  \Description{seq length experiment}
\end{figure}

An analysis of Figure~\ref{fig:seq_len} reveals that there is a positive correlation between sequence length and performance. The performance improvement increases at a rate that is sub-linear with respect to the sequence length. The use of concatenation as the early fusion method was found to be superior to the use of appending. Therefore, the optimal engagement gain can be achieved by utilizing the maximum available sequence length and employing concatenation as the early fusion method.

\subsubsection{Transformer hyperparameters}
\begin{figure}[h]
  \centering
  \includegraphics[width=\linewidth]{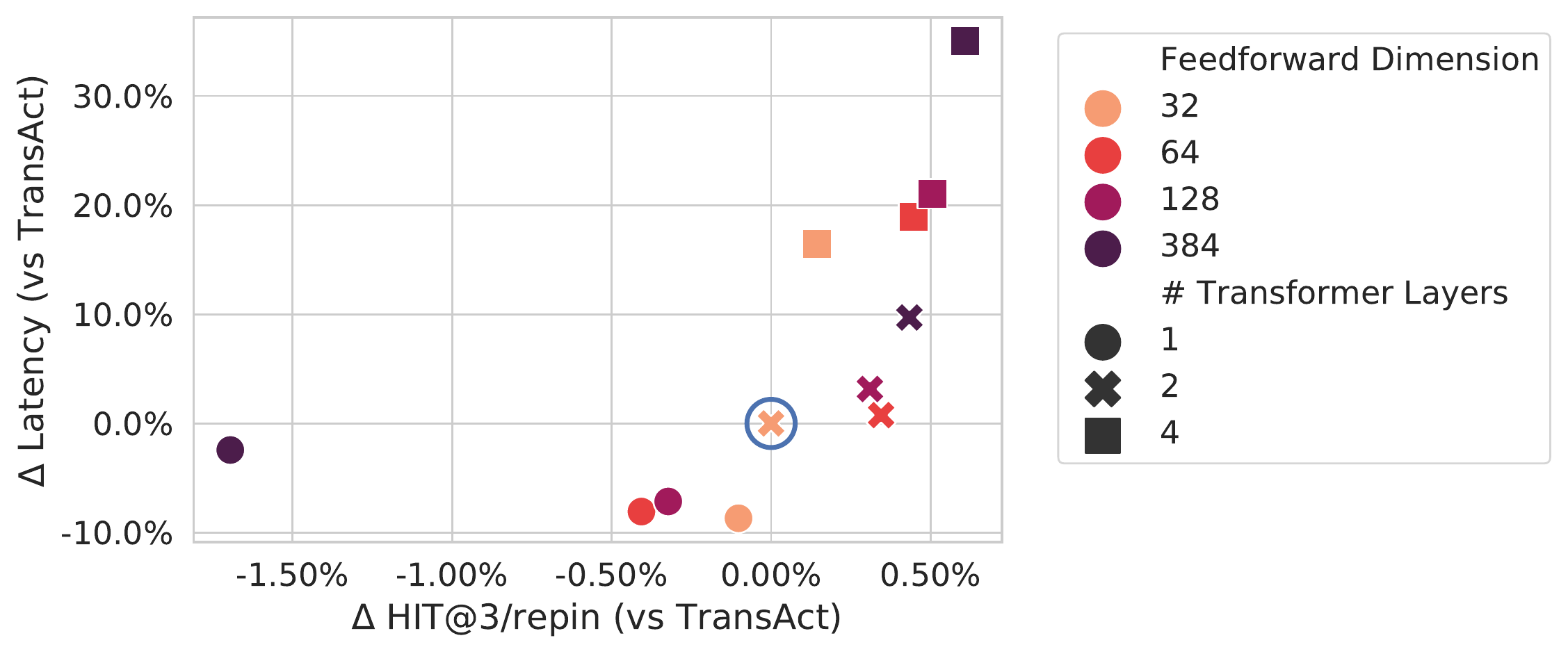}
  \caption{Effect of transformer hyperparameters on model performance and latency}
  \label{fig:tfmr_hpt}
  \Description{seq length experiment}
\end{figure}

We optimized TransAct's transformer encoder by adjusting its hyperparameters. As shown in Figure~\ref{fig:tfmr_hpt},
increasing the number of transformer layers and feed forward dimension leads to higher latency and also better performance.
While the best performance was achieved using 4 transformer layers and 384 as the feed forward dimension, this came at the cost of a 30\% increase in latency, which does not meet the latency requirement.
To balance performance and user experience, we chose 2 transformer layers and 32 as the hidden dimension.

\subsubsection{Transformer output compression}
\label{sec:compression}
The transformer encoder produces $\vO \in \mathbb{R}^{d \times |S|}$, with each column corresponding to an input user action. 
However, directly using $\vO$ as input to the DCN v2 layers for feature crossing would result in excessive time complexity, which is quadratic to the input size.

To address this issue, we explored several approaches to compress the transformer output.
Figure~\ref{fig:seq_encoder} shows that the highest HIT@3/repin is achieved by combining the first K columns and applying max pooling to the entire sequence. 
The first K column represents the most recently engaged pins and the max pooling is an aggregated representation of the entire sequence,
Although using all columns improved HIT@3/hide slightly, the combination of the first K columns and max pooling provided a good balance between performance and latency. We use K=10 for TransAct.

\begin{table}
  \caption{Ablation study of transformer output compression}
  \label{tab:tfmr_output}
\resizebox{.45\textwidth}{!}{%
\begin{tabular}{ccrr}
    \toprule
    Output Compression &  Size & HIT@3/repin &  HIT@3/hide   \\
    \midrule
        a random  col    & $d$ &+6.80\%  & -10.96\%   \\
        first  col       & $d$ &+7.82\%     & -11.28\%  \\
    random K  cols     & $Kd$&  +7.42\%     & -12.12\%       \\
    first K cols    & $Kd$ & +9.38\%   & -14.33\% \\
    all cols    & $|S|d$&+8.86\%   & \textbf{-15.70\%}  \\
    max pooling      & $d$ &+6.38\%   & -14.15\% \\
    \textbf{first K cols + max pool} & $(K+1)d$ &\textbf{+9.41\%}  & -14.86\% \\
    all cols + max pool   & $(|S|+1)d$ &+8.67\%  & -12.64\% \\
  \bottomrule
\end{tabular}%
}
\end{table}


\subsection{Online Experiment}
Compared with offline evaluation, one advantage of online experiments in recommendation tasks is that they can be run on live user data, allowing the model to be tested in a more realistic and dynamic environment. 
For the online experiment, we serve the ranking model trained on the 2-week offline training dataset. We set the control group to be the Pinnability model without any realtime user sequence features. The treatment group is Pinnability model with TransAct. Each experiment group serves 1.5\% of the total users who visit Homefeed page.

\subsubsection{metrics}
On Homefeed, one of the most important metrics is \textbf{Homefeed repin volume}. Repin is the strongest indicator that users find the recommended pins relevant, and is usually positively correlated to the amount of time users spend on Pinterest. Empirically, we found that offline HIT@3/repin usually aligns very well with Homefeed online repin volume. Another important metric is \textbf{Homefeed hide volume}, which measures the proportion of recommended items that users choose to hide or remove from their recommendations. High hide rates indicate that the system is recommending items that users do not find relevant, which can lead to a poor user experience. Conversely, low hide rates indicate that the system is recommending items that users find relevant and engaging, which can lead to a better user experience.

\subsubsection{Online engagement}

We observe significant online metric improvement with TransAct introduced to ranking. Figure~\ref{tab:online_metric} shows that we improved the Homefeed repin volume by 11\%.  It’s worth noting that engagement gains for non-core users are higher because they do not have a well-established user action history. And realtime features can capture their interest in a short time. Using TransAct, the Homefeed page is able to respond quickly and adjust the ranking results timely. We see hide volume dropped and that the overall time spent on Pinterest is increased.
\begin{table}
  \caption{Online evaluation of TransAct }
  \label{tab:online_metric}
  \begin{tabular}{crr}
    \toprule
    Online Metrics & All Users & Non-core Users\\
    \midrule
    Homefeed Repin Volume & +11.0\% & +17.0\%\\
    Homefeed Hide Volume & -10.0\% & -10.5\%\\
    Overall Time Spent & +2.0\% & +1.5\%\\
  \bottomrule
\end{tabular}
\end{table}

\subsubsection{Model retrain}
\label{sec:retrain_exp}
\begin{figure}[h]
  \centering
  \includegraphics[width=0.8\linewidth]{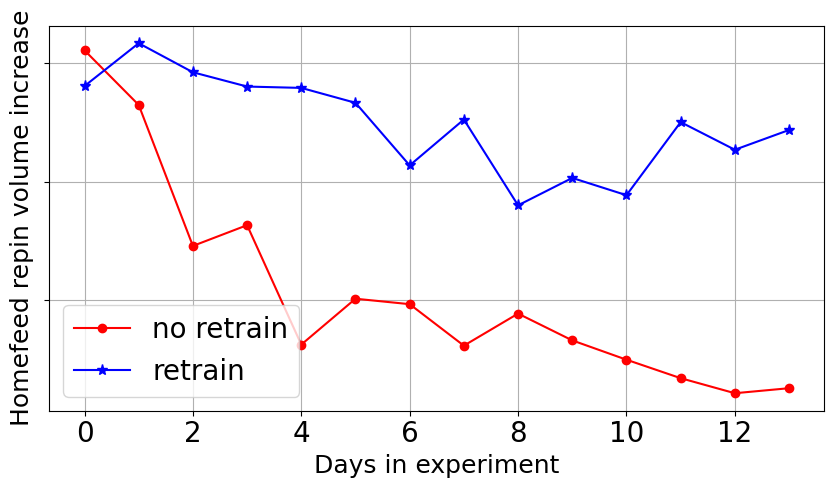}
  \caption{Effect of retraining on TransAct}
  \label{fig:retrain1}
\end{figure}

One challenge observed in the TransAct group was the decay of engagement metrics over time for a given user.
As shown in Figure~\ref{fig:retrain1}, we compare the Homefeed repin volume gain of TransAct to the baseline, with both groups either fixed or retrained.
We observed that if TransAct was not retrained, despite having a significantly higher engagement on the first day of the experiment, it gradually decreased to a lower level over the course of two weeks. 
However, when TransAct was retrained on fresh data, there was a noticeable increase in engagement compared to not retraining the model. 
This suggests that TransAct, which utilizes realtime features, is highly sensitive to changes in user behavior and requires frequent retraining. 
Therefore, it is desired to have a high retrain frequency when using TransAct. In our production, we set the retrain frequency to twice a week and this retrain frequency has been proven to keep the engagement rate stable.

\subsubsection{Random time window masking}

Another challenge observed was dropping diversity in recommendations. Diversity measures the broadness and variety of the items being recommended to a user. Previous literature\cite{gg_surrogate} finds diversity is associated with increasing user visiting frequency. However, diversity is not always desirable as it can lead to a drop in relevance. Therefore, it is crucial to find the right balance between relevance and diversity in recommendations.

At Pinterest, we have a 28k-node hierarchical interest taxonomy~\cite{interest_blog} that classifies all the pins. The top-level interests are coarse. Some examples of top-level interests are art, beauty, and sport. Here, we measure the \textbf{impression diversity} as the summation of the number of unique top-level interests viewed per user. We observe that with TransAct introduced to Homefeed ranking, the impression diversity dropped by 2\% to 3\%. The interpretation is that by adding the user action sequence feature, the ranking model learns to optimize for the user’s short-term interest. And by focusing on mainly short-term interest, the diversity of the recommendation dropped. 

We mitigate the diversity drop by using a random time window mask in the transformer as mentioned in Section~\ref{sec: seq_model}. This random masking encourages the model to focus on content other than only the most recent items a user engaged with. With this design, the diversity metric drop was brought back to only -1\% without influencing relevance metrics like repin volume. 
We also tried using a higher dropout rate in the transformer encoder layer and randomly masking out a fixed percentage of actions in the user action sequence input. However, neither of these methods yielded better results than using random time window masking. They increased the diversity at the cost of engagement drop. 

\section{Discussion}
\label{sec:discussion}
\subsection{Feedback Loop}


An interesting finding from our online experiment is that the true potential of TransAct is not fully captured. We observed a greater improvement in performance when the model was deployed as the production Homefeed ranking model for full traffic. This is due to the effect of a positive feedback loop: as users experience a more responsive Homefeed built on TransAct, they tend to engage with more relevant content, leading to changes in their behavior (such as more clicks or repins). These changes in behavior lead to shifts in the realtime user sequence feature, which are then used to generate new training data. Retraining the Homefeed ranking model with this updated data results in a positive compounding effect, leading to a higher engagement rate and a stronger feedback loop. This phenomenon is similar to "direct feedback loops" in literature~\cite{tech_debt} which refers to a model that directly influences the selection of its own future training data, and it is more difficult to detect if they occur gradually over time.


\subsection{TransAct in Other Tasks}
The versatility of TransAct extends beyond just ranking tasks. It has been successfully applied in the contextual recommendation and search ranking scenarios as well. 
TransAct is used in \textbf{Related Pins}~\cite{related_pins} ranking, a contextual recommendation model to provide personalized recommendations of pins based on a given query pin. 
TransAct is also applied in Pinterest's \textbf{Search} ranking~\cite{search_blog} system and \textbf{notification} ranking~\cite{pins_notif}. Table~\ref{tab:other_app} showcases the effectiveness of TransAct in a variety of use cases and its potential to drive engagement in more real-world applications.

\begin{table}
  \caption{TransAct's impact on other applications}
  \label{tab:other_app}
  \begin{tabular}{ccc}
    \toprule
    Application & Metrics & $\Delta$\\
    \midrule
    Related Pins & Repin Volume & +2.8\%\\
    \midrule
     Search  & Repin Volume & +2.3\%\\
     \midrule
    \multirow{2}{*}{Notification }  & Email CTR &+1.4\%  \\
    & Push Open Rate &+1.9\% \\
  \bottomrule
\end{tabular}%
\end{table}

\section{Conclusions}
\label{sec:conclusion}
In this paper, we present TransAct,  a transformer-based realtime user action model that effectively captures users' short-term interests by encoding their realtime actions.
Our novel hybrid ranking model merges the strengths of both realtime and batch approaches of encoding user actions, and has been successfully deployed in the Homefeed recommendation system at Pinterest. 
The results of our offline experiments indicate that TransAct significantly outperforms state-of-the-art recommender system baselines.  
 In addition, we have discussed and provided solutions for the challenges faced during online experimentation, such as high serving complexity, diversity decrease, and engagement decay. 
 The versatility and effectiveness of TransAct make it applicable for other tasks, such as contextual recommendations and search ranking.


\bibliographystyle{ACM-Reference-Format}
\balance
\bibliography{main}

\appendix

\section{Head Weighting}\label{appendix:head_weights}

We illustrate how head weighting helps the multi-task prediction task here. Consider the following example, of a model using 3 actions: repins, clicks, and hides. The label weight matrix is set as Table~\ref{tab:labelweight}.

\begin{table}[!ht]
  \caption{An example of label weight matrix $\mM$ with 3 actions}
  \label{tab:labelweight}
  \begin{tabular}{|c|ccc|}
  \hline
    \diagbox{Head}{Action} &click & repin  & hide\\
    \hline
    click & 100 & 0 & 100\\
    repin & 0 & 100 & 100 \\
    hide & 1 & 5 & 10\\
  \hline
\end{tabular}
\end{table}

Hides are a strong negative action, while repins and clicks are both positive engagements, although repins are considered a stronger positive signal than clicks.
We set the value of $\mM$ manually, to control the weight on cross-entropy loss. Here, we give some examples of how this is achieved.
\begin{itemize}

\item If a user only hides a pin ($\vy_{hide}=1$), and does not repin or click ($\vy_{repin}=\vy_{click}=0$). Then we want to penalize the model more if it predicts repin or click, by setting the $\mM_{repin, hide}$ and $\mM_{click, hide}$ to a large value. 


$w_{repin}=\mM_{repin, hide} * \vy_{hide} =100$.

$w_{click}=\mM_{click, hide}  * \vy_{hide}  =100$.

\item If a user only repins a pin ($\vy_{repin}=1$), but does not hide or click ($\vy_{hide}=\vy_{click}=0$). We want to penalize the model if it predicts hide: $w_{hide} = \mM_{hide,repin} * \vy_{repin} = 5$. 

But we do not need to penalize the model if it predicts a click, because a user could repin and click the same pin.
$w_{click} =  \mM_{click, repin} * \vy_{repin} = 0$
 \end{itemize}




\section{Positional Encoding}
We tried several positional encoding approaches: learning positional embedding from scratch, sinusoidal positional encoding~\cite{tfmr}, and linear projection positional encoding as proposed in~\cite{alibaba_seq_tfmr}.
Table\ref{tab:pe} shows that positional encoding does not add much value.
\label{appendix:pe}
\begin{table}[!ht]
  \caption{Offline evaluation of different positional encoding methods compared with TransAct}
  \label{tab:pe}
  \begin{tabular}{crr}
    \toprule
    Positional encoding method &  HIT@3/hide &  HIT@3/repin  \\
    \midrule
    None (TransAct)      &  -      & -\\
    From scratch & +0.86\%   & -0.61\%\\
    Sinusoidal  & +0.78\%     &-0.13\%\\
    Linear projection $^*$  & +2.29\%       &   +0.19\%  \\
  \bottomrule
\end{tabular}
\end{table}

\section{Model Efficiency}\label{appendix:efficiency}

Table~\ref{tab:efficiency} shows more detailed information on the efficiency of our model, including number of flops, model forward latency per batch (batch size = 256), and serving cost. The serving cost is not linearly correlated with model forward latency because it is also related to server configurations such as time out limit, batch size, etc. GPU serving optimization is important to maintain low latency and serving cost. 
\begin{table}[!ht]
  \caption{Model Efficiency Numbers from Serving Optimization}
  \label{tab:efficiency}
  \begin{tabular}{cccc}
    \toprule
     &  Baseline(CPU) &  TransAct(CPU) & TransAct(GPU)  \\
    \midrule
    Parameters      & 60M  & 92M & 92M\\
    flops & 1M   & 77M & 77M\\
    Latency  & 22ms     & 712ms & 8ms\\
    Serving Cost  & 1x       &   32x & 1x  \\
  \bottomrule
\end{tabular}
\end{table}

\end{document}